# THE ORIGIN OF ENERGY FOR THE BIG BANG

Xinyong Fu, Zitao Fu, Shanghai Jiao Tong University, Email: xyfu@sjtu.edu.cn,

Address: Room302, Number 7, Lane 24, South Yili Road, Shanghai, 201103 China P. R.

**Abstract**

Our universe is probably a huge black hole. If that is true, all the light and heat ejected by various celestial bodies into the space will be confined within it and shuttling ceaselessly, leading eventually to a uniform equilibrium radiation at certain temperature. The authors hold that the 3 K background radiation discovered in 1965 is actually such equilibrium radiation. The 3 K background radiation is convincible evidence that our universe is a closed one and it is actually a huge heat ocean.

Billions of galaxies produced in the big bang will also be shuttling within the closed universe, passing the central part of the universe again and again. They have many chances to meet each other. There are numerous black holes of various sizes in these galaxies. A black hole absorbs matter and radiation, even swallows in other approaching celestial bodies. These black holes provide the mechanism of matter-re-gathering and energy-re-gathering in the universe. All the real matter will finally gather again to form a single huge black hole. The huge black hole absorbs energy from the heat ocean further, until its energy overpasses certain threshold value, leading to a new big bang.

The heat ocean has energy, hence it has mass, too. Calculations show that, the mass of the heat ocean surpasses overwhelmingly the mass of all the real matter. The heat ocean is the dominant part of the universe, and the real matter is only a fraction, which explodes and re-gathers repeatedly within the universe.

## 1. A fundamental problem in cosmology

The idea that the present universe----billions of galaxies----was produced in a big bang about 13 billion years ago is approved of by most of today's astrophysicists. Many of them tend to hold further that after the big bang and expansion, there will be a contraction, and all the matter will gather again, and then a new big bang, and so on.

In all these arguments, however, there is a problem, which is obviously very fundamental, but so many people turn a blind eye to it: for every big bang, a great amount of energy is needed, but where does it come from?

Let us look at the problem more closely.

It seems as if everyone knows where this energy will go. After a big bang, the energy of the explosion converts into kinetic energy, gravitational potential energy, nuclear energy, and so on. Although these different forms of energy may convert each other in many later processes, they will all inevitably go to a same final destination: to be changed into light and heat and be radiated into the vast cosmic space. Let's see some examples. When nebulae condense to form stars, a great amount of gravitational potential energy converts into light and heat, which are then radiated into the vast cosmic space. After stars are formed, during their about ten-billion-year lifetimes, huge amount of nuclear energy of hydrogen converts into light and heat, which are also radiated into the vast cosmic space. And, wherever collisions or frictions occur, kinetic energy converts into light and heat, which, too, are radiated into the vast cosmic space. And so on.

Then, after entering the space, what will happen to these light and heat? Where will they finally go? What's their future fate?

It seems as if nobody in the circles of physics and astrophysics today is interested in this problem. In some current theories, after the big bang and expansion, all the matter will contract and get gathered again, preparing for a new big bang. However, never the escaped light and heat are mentioned to come back again.

Actually, what we meet here is a taboo in the present thermodynamics. According to Clausius' second law, all the spontaneous processes mentioned above (energy of other forms converts into light and heat and then scatters into the vast space) result in the increase of entropy, and they are all irreversible. It is widely accepted that the second law is an "impossible barrier" to the re-concentration of the energy that has scattered into the vast cosmic space.

Thus, with every big bang and the consequent expansion and contraction, the matter of the universe will lose a great amount of energy. Never will the energy available go exhausted? How can the universe explode again and again?

This is certainly a fundamental problem in cosmology. The simplest mathematic logic reveals that we should not evade such a problem on any account.

## 2. A heat ocean within a closed universe

The general relativity introduced a discussion about gravitationally closed celestial bodies. Whether a celestial body is closed, or, whether it is a black hole, is determined by $R = 2 \, GM / c^2$. That is, how much mass there is in how small a volume. A black hole may be as large as a star, or as a galaxy, even as the universe as a whole, see Table 1 [1].

| Size | $M / M_\odot$ | $\rho$ (kg/m$^3$) | $R = 2GM / c^2$ |
|---|---|---|---|
| star | 1 | 2 x 10$^{19}$ | 3 km |
| galaxy | 10$^{11}$ | 2 x 10$^{-3}$ | 0.1 light year |
| "universe" | 10$^{22}$ | 2 x 10$^{-25}$ | 10$^{10}$ light year |
| 2.74K heat ocean | 1.5×10$^{24}$ | 5 x 10$^{-31}$ | 2 x 10$^{12}$ light year |

Table 1. Main parameters of black holes of different sizes. The data in the first three cases are quoted from E. H. Avrett *Frontiers of Astrophysics*. The fourth case is developed by the authors.

Is our universe really a closed one? This depends on the amount of its mass and mass distribution. According to the current opinion, the relevant data obtained so far is still inadequate, so we cannot yet arrive at a definite conclusion. Nevertheless, a part of the astrophysicists today tend to hold that, very probably, our universe is a huge black hole.

If that is true, what will the things be?

First of all, all the light and heat ejected into the vast space will be confined within the closed universe. Every piece of the radiation ejected into the space tends to travel along its present direction, and will finally approach the edge of the closed universe, namely, the 'event horizon". It will be drawn back and then go toward the central part of the universe. All the radiation will, in such a way, be shuttling on end within the Schwarzschild sphere of the universe. According to the basic ideas of thermodynamics, no matter what the initial situation is, after a sufficiently long period of time, all these radiations will gradually reach a common equilibrium state, and, a common temperature. The eventual situation is very similar to the one of the equilibrium radiation (the blackbody radiation) within the cavity of a solid at a constant temperature.

As is well discussed in thermodynamics, the equilibrium radiation within a cavity at a constant temperature has three typical characteristics: (1) it has a spectrum that coincides with Planck's formula, (2) an ideal isotropy, (3) an ideal stableness with respect to time. The equilibrium radiation within the closed universe should also have such three typical characteristics.

In 1965, Penzios and Wilson discovered accidentally the 3 K microwave background radiation coming from the remote space. This background radiation shows three characteristics: its spectrum coincides with Planck's formula, and it shows an ideal isotropy as well as an ideal stableness [2]. They are exactly the same three characteristics of an equilibrium radiation. Accordingly, the authors hold that, the 3 K microwave background radiation is actually just the above mentioned equilibrium radiation within the closed universe.

The microwave background radiation discovered by Penzios and Wilson is convincible evidence that our universe is a closed one.

So now, our universe is a huge black hole. Its interior is filled with an equilibrium radiation at a certain temperature, to be more accurate, $T$ = 2.74 K. The closed universe is actually a vast heat ocean. All the light, heat and other electromagnetic radiations ejected by all the celestial bodies, and even the kinetic energy of the high-speed particles in cosmic rays, and so on, will merge into this heat ocean to reach a common equilibrium state. The vast and almost transparent universe demands a very long relaxation time for these radiations and particle motions to reach a common equilibrium state. Nevertheless, as the universe is closed, whatever long relaxation time is available.

Where heat flows to, entropy follows. The vast heat ocean is also an ocean of entropy.

## 3. Matter gathering and energy gathering

Due to the big bang, all the real matter in our universe, i.e., billions of galaxies, is now flying outward in all the directions at various high speeds [3]. As the universe is a huge black hole, all these galaxies, sooner or later, will slow down and then turn back toward the central part of the universe. They will also be shuttling within the universe, passing through the central part repeatedly.

There are numerous and numerous celestial bodies in all these galaxies, including numerous black holes of different sizes. Every black hole, from the moment of its formation, starts to absorb ceaselessly all the matter and radiation approaching it, and hence gets larger. A black hole can even swallow in other celestial bodies, including other black holes. As the galaxies shuttling in the universe, passing through the central part repeatedly, they have chances to meet each other again and again. There will be, among all the black holes and other celestial bodies, numerous approaches, collisions, annexations, and so on. Hence, the black holes will inevitably get larger and larger, while their number gets smaller and smaller. The larger a black hole is, the more chance it has to take in other objects. After a very long period of time, there should be only a small number of large black holes left in the universe, and the annexation process becomes faster. Finally, only one is left. Of course, this is a huge one, being very compact, and containing almost all the real matter of the universe.

So, it is the black holes that provide the mechanism of gathering the matter again in the universe.

Now, the picture of the universe becomes rather simple. The whole universe is a very huge and dilute black hole, actually, a vast heat ocean. In the central part of the heat ocean, there is another black hole, which is relatively small, containing almost all the real matter of the universe.

Besides gathering real matter, the black holes have another important function in the universe-----gathering energy again.

As mentioned above, all the black holes absorb radiation from the heat ocean ceaselessly during their lifetimes. The authors hold that these processes reduce the entropy of the universe. Stars give off light and heat into the space, resulting in the increase of entropy; black holes, quite the contrary, take in the radiation from the space (from the heat ocean), resulting in the decrease of entropy. The two kinds of processes are obviously opposite to each other.

Numerous black holes in the universe will gather up a very great amount of energy in their lifetimes. When they finally combine to form a huge central black hole, all the heat collected by them in the previous stages will be brought into it.

Now let's look at the final huge black hole. Once formed, it also starts to collect radiation from the heat ocean ceaselessly. As it has a very large Schwarzschield sphere, it does the work at a rather high rate. Its energy increases quickly and monotonically. After a very long period, the extremely high concentration of energy within it will turn its interior matter into some extraordinary state. The author refers to the matter in such an extraordinary state as the pre-hydrogen substance. Pre-hydrogen substance should actually be the "primordial matter" or "ylem" in the big bang theory.

As the central black hole continues to absorb thermal radiation, its interior energy increases further. We cannot imagine or accept that such a process will go on without an end. Once the interior energy overpasses some threshold value, a new big bang breaks out. Large amount of the pre-hydrogen substance changes into hydrogen and helium and flies outward at extremely high speed into the vast space. All the thermal radiation collected by the black holes will once again change into kinetic energy, gravitational potential energy, as well as the nuclear energy of hydrogen and helium, and so on.

Black holes can prevail against the "impassable barrier" of the second law of thermodynamics. They re-gather and refresh the energy in the universe.

The central black hole can be referred to as the "egg" of the universe. It takes a very long time for the heat ocean to feed the "egg" to grow up, as the energy density of the equilibrium radiation of the heat ocean is low.

## 4. Main parameters of the heat ocean

Our universe is, as discussed above, a black hole. Its radius is determined by $R = 2GM/c^2$, where $M$ stands for its total mass, which includes two parts:

$$M = M_1 + M_2 \ .$$

$M_1$ represents the mass of all the real matter, including the mass of several billions of galaxies, the "dark substance", and so on. Approximately we have

$$M_1 \approx 5 \times 10^9 \times 10^{11} M_\odot \times 10 = 5 \times 10^{21} M_\odot \ ;$$

$M_2$ represents the mass of the vast heat ocean, i.e., the mass of the 3 K equilibrium radiation in the universe. The authors' opinion is that $M_2$ should be much larger than $M_1$, which is explained as follows.

When all the real matter $M_1$ is concentrated in the narrow space of the "egg', the gravitational actions between its different parts is terribly strong. Hence the amount of energy needed to produce a big bang, $\Delta E$ (absolute value), should be extraordinarily large.

Then, how large should this amount of energy be? A fundamental and rough estimation is: $\Delta E$ should be much larger than $M_1 c^2$, the rest-mass energy of $M_1$.

Why?

Just after the big bang, all the real matter ($M_1$) rushes outward in all directions at speeds close to light speed. $M_1$ is divided into many small parts, with $m_i$ being the rest mass of the $i$-th part, which rushes outward at a speed of $v_i$. The actual mass of $m_i$ at that time is $m_i'$, which can be expressed as

$$m_i' = \frac{m_i}{\sqrt{1 - v_i^2 / c^2}} \ ,$$

The total actual mass at that time is then

$$M_1' = \sum m_i' = \sum \frac{m_i}{\sqrt{1 - v_i^2 / c^2}} \ ,$$

Introducing in $\overline{v^2}$ as the average square speed for all the various speeds of the different parts in the above expression, we can rewrite $M_1'$ as follows

$$M_1' = \sum \frac{m_i}{\sqrt{1 - \overline{v^2} / c^2}} = \frac{M_1}{\sqrt{1 - \overline{v^2} / c^2}} \ ,$$

As most of the various speeds $v_i$ are very high, close to light speed, so $\sqrt{\overline{v^2}}$ is close to light speed, too. Hence, $M_1^{'}$ is much larger than $M_1$, and we have

$$M_1^{'}c^2 \gg M_1 c^2 ,$$

and

$$M_1^{'}c^2 - M_1 c^2 \approx M_1^{'}c^2 .$$

The increase of energy from $M_1 c^2$ to $M_1^{'} c^2$ results from the black holes' absorbing and accumulating radiation from the heat ocean. When a big bang occurs, besides $M_1^{'}c^2$, there is also a great amount of energy released in the form of heat radiation at extremely high temperature. This amount of energy also results from the black holes' absorbing and accumulating radiation from the heat ocean. Hence, $\Delta E$, the total energy drawn from the heat ocean by the black holes to result in eventually a big bang, should be much greater than $M_1^{'}c^2 - M_1 c^2$, or briefly, much greater than $M_1^{'}c^2$. So we have

$$\Delta E > M_1^{'}c^2 \gg M_1 c^2 ,$$

Let us proceed further. The heat ocean is actually an energy reservoir from which the black holes (include the "egg") draw sufficient energy for a future big bang. Hence, the amount of the energy of the heat ocean, $E$, must be much larger than $\Delta E$. We then have

$$E \gg \Delta E > M_1^{'}c^2 \gg M_1 c^2 ,$$

Dividing through by $c^2$, we get the mass of the heat ocean

$$M_2 = \frac{E}{c^2} \gg \frac{\Delta E}{c^2} > M_1^{'} \gg M_1 ,$$

i.e.

$$M_2 \gg M_1 .$$

So, we come to a conclusion that the mass of the heat ocean, $M_2$, is much larger than the mass of all the real matter, $M_1$. Compared to the mass of the heat ocean, the mass of all the real matter can be neglected

$$M = M_1 + M_2 \approx M_2 \ .$$

So, if we now use $R = 2GM/c^2$ to calculate the Schwarzschild radius of the universe, we may replace $M$ by $M_2$. That means, we can consider that the universe is being gravitationally closed only due to the mass of the heat ocean. Such a model is easy to deal with, and the calculations will be simple and accurate. Based on such a model, we calculate the main parameters of the heat ocean as follows.

### (1) Mass density of the heat ocean

According to Stefan-Boltzmann's law, the radiation intensity of the equilibrium radiation at temperature $T$=2.74K is

$$\Gamma = \sigma T^4 = 3.20 \times 10^{-6} \text{ J m}^{-2}\text{s}^{-1} ,$$

On the other hand, $\Gamma = \frac{c}{4} u = \frac{c}{4}(\alpha T^4)$, hence the energy density of the equilibrium radiation of the heat ocean is

$$u = \alpha T^4 = 4\Gamma/c = 4.26 \times 10^{-14} \text{ J m}^{-3} ,$$

The corresponding mass density of the equilibrium radiation is then

$$\rho = \frac{u}{c^2} = 4.13 \times 10^{-31} \text{ kg m}^{-3} ,$$

### (2) The radius and total mass of the heat ocean

As the universe can be considered being closed due to the mass of the heat ocean alone (the mass of the real matter can be neglected), we have

$$R = \frac{2GM}{c^2} \approx \frac{2GM_2}{c^2} = \frac{2G}{c^2}(\frac{4}{3}\pi R^3 \rho) \quad ,$$

Solving for $R$, we have

$$R = \sqrt{\frac{3c^2}{8\pi G \rho}} = 1.85 \times 10^{28} \text{ m} = 1.95 \times 10^{12} \text{ light years} \quad ,$$

R is about 2000 billion light years.

The total mass of the heat ocean is
$$M_2 = \frac{4}{3}\pi R^3 \rho = 3.13 \times 10^{54} \text{ kg} = 1.57 \times 10^{24} M_\odot,$$

The total mass of the heat ocean is about 300 times of all the real matter in the universe ($M_1 \approx 5 \times 10^{21} M_\odot$).

So, the main body of our universe is really the vast heat ocean, all the real matter is only a small fraction, which explodes and gathers repeatedly within the heat ocean.

Now, let's suppose that during a very long period of time, the "egg" and the previous black holes have absorbed totally one tenth of the total energy of the heat ocean,
$$\Delta E = 0.1 E = 0.1(M_2 c^2) = 0.1(300 M_1 c^2) = 30 M_1 c^2 \quad .$$

This amount of energy, 30 $M_1 c^2$, is very large for the "egg", as the original rest mass of the real matter in the "egg" is only $M_1$. It may be already enough for the "egg" to start a big bang. Calculations show that corresponding to such a reduce in its energy, the heat ocean will cool down from 2.74 K to 2.67 K. We see that the temperature of the heat ocean is rather stable.

If $\Delta E = 30\ M_1 c^2$ is still not enough to start a big bang, then let it be 60 $M_1 c^2$, etc.

## 5. Concluding remarks

On the surface of the earth, the vast ocean plays a dominant role in water cycling. It is the final destination of almost all the rivers, meanwhile, it is also the origin of the water for these rivers.

Similarly, in the closed universe, the 3 K background radiation, i.e., the vast heat ocean, plays a dominant role in energy cycling. It is the final destination of all the light, heat and other energies ejected by all the celestial bodies during the big bang and the subsequent processes, meanwhile, it is also the origin of the energy for the big bangs.

Matter cycles, energy cycles, the universe will never come to an end of "heat death."